\documentstyle[
               12pt,%
               prd,aps]{revtex}

\newcommand{\const}{\mbox{const} }

\begin{document}
\title{More about Vacuum Spacetimes with Toroidal Null Infinities} 

\author{{\bf Peter H\"ubner}\\
        (pth@aei-potsdam.mpg.de)\\
        Max-Planck-Institut f\"ur Gravitationsphysik\\
        Albert-Einstein-Institut\\
        Schlaatzweg 1\\
        D-14473 Potsdam\\
        FRG}

\maketitle
\thispagestyle{empty}

\begin{abstract}
Recently Bernd Schmidt has given three explicit examples of spacetimes with
toroidal null infinities. In this paper all solutions with a toroidal
null infinity within Schmidt's metric ansatz (polarized Gowdy models) are
constructed. The members of the family are determined by two smooth
functions of one variable. For the unpolarized Gowdy models the same
kind of analysis carries through. 
\end{abstract}
\section{Introduction}
In \cite{Sc96vs} Bernd Schmidt gave three examples of spacetimes which
can be mapped to an unphysical\footnote{See e.~g.\ \cite[Definition~1
  and the following text]{Hu95gr} for an explanation of the
  distinction between physical and unphysical spacetimes in the context of
  asymptotical flatness.} spacetime with regular boundary with the topology
``torus times line''. Those
examples fulfill the conditions regarding regularity in the definition of
asymptotic simplicity, but not the condition that every null
geodesic starts at past null infinity and ends at future null
infinity\footnote{With this condition the conformal boundary must have
  the topology ``sphere times line'' \cite{Pe64ct}}. 
As the metric A3 in the
classification of Ehlers and Kundt is the simplest known solution with
this structure of null infinity I will call those solutions
\textbf{asymptotically A3}.
\\
If one solves an initial value problem numerically in physical
spacetime, the topological differences of asymptotically A3 solutions to
asymptotically flat solutions do not show up in a essential way.
The grid has periodic boundary conditions with respect to two
spacelike coordinates, and ``normal'' boundary conditions for the
remaining spacelike coordinate.  Gravitational radiation in
asymptotically A3 spacetimes is emitted
through the boundaries limiting the range of the latter coordinate. 
Calculating the gravitational radiation emitted is pure algebra
(see equation (\ref{mBondi}) and \cite{Fo97XX}). Therefore these
solutions may not only serve as 
testbeds for the numerical calculation of solutions from initial data
(especially for the outer boundary treatment) but they may also
serve as testbed for the radiation extraction procedures.
\\
The explicit solutions given by Bernd Schmidt originate in the
solution of a 1+1 wave equation. For any smooth data (two functions of
one variable) one gets a smooth solution of Einstein's equation, but
not necessarily one which is asymptotically A3. He conjectured
that every solution of the 1+1 wave equation which can be extended
smoothly to null infinity yields an asymptotically A3 solution, i.~e.\
the other function determining the metric necessarily has a smooth
limit at null infinity also.
\\
In this article I give necessary and sufficient conditions for
asymptotically A3 solutions on the initial data of the wave
equation(s) for polarized and unpolarized Gowdy spacetimes. For
polarized Gowdy models further explicit solutions are
guessed. Generally they radiate away part of their Bondi mass.
\section{Construction of the solutions}
As described in \cite{Sc96vs}\footnote{Note that the metric is written
  in terms of the functions $\tilde W$ and $\tilde M= - 1/4 \>\ln \tilde
  t + \tilde N/2$ in reference~\cite{Sc96vs}.}, for the metric 
\begin{equation}
  \label{physMetrik}
  \tilde g = \frac{1}{\sqrt{\tilde t}} \> e^{\tilde N} 
               ( -d{\tilde t}^2 + d{\tilde z}^2 ) 
             + \tilde t \, ( e^{\tilde W} d{\tilde x}^2 
                             + e^{-\tilde W} d{\tilde y}^2 ) 
\end{equation}
the field equations are
\begin{mathletters}
\label{physGl}
\begin{eqnarray}
  \tilde W_{,\tilde t\tilde t} 
  + \frac{1}{\tilde t} \, \tilde W_{\tilde t}
  - \tilde W_{,\tilde z\tilde z}
  & = & 0, \label{physWllnGlW} \\
  \tilde N_{,\tilde t}
  - \frac{\tilde t}{2} 
  \left( \left( \tilde W_{,\tilde t} \right)^2
         + \left( \tilde W_{,\tilde z} \right)^2 \right)
  & = & 0, \label{phystGlM} \\
\noalign{and}
  \tilde N_{,\tilde z} - \tilde W_{,\tilde t} \, \tilde W_{,\tilde z}
  & = & 0. \label{physrGlM}
\end{eqnarray}
\end{mathletters}
$\tilde W$ and $\tilde N$ are functions of $\tilde t$ and $\tilde z$.
The $\tilde{\hphantom{M}}$ flags quantities which are
supposed to live in physical spacetime in contrast to objects without
$\tilde{\hphantom{M}}$ which live in unphysical spacetime.
The wave equation (\ref{physWllnGlW}) is the integrability condition
of the equations (\ref{phystGlM}) and (\ref{physrGlM}). Therefore,
giving smooth data $\tilde W$ and $\tilde W_{,\tilde t}$ on any
spacelike surface with $\tilde t = \tilde t_0$, one obtains smooth
functions $\tilde W(\tilde t,\tilde z)$ and $\tilde N(\tilde t,\tilde
z)$ for all $\tilde t>0$. But those are not, depending on the data,
necessarily asymptotically A3.
\\
To show that such a spacetime $(\tilde M,\tilde g)$ is indeed
asymptotically A3 one has to find a conformal mapping $\Omega$
with $g = \Omega^2 \tilde g$ and an unphysical spacetime $(M,g)$ with
smooth boundary ${\cal J}$. 
To find all asymptotically A3 solutions it is advantageous to proceed in the
other direction, i.~e.\ to find equations for the functions $W$ and
$N$ in the unphysical spacetime which guarantee that (\ref{physGl}) is
fulfilled for the corresponding $\tilde W$ and $\tilde N$ and which
ensure that for smooth initial data for $W$ and $N$ on $M$ the
functions $W$ and $N$ are smooth in a sufficient region of $M$. The
functions $\tilde W$ and $\tilde N$ on physical spacetime are related
to $W$ and $N$ on unphysical spacetime by ${\tilde W(\tilde t,\tilde
  z) = W(t(\tilde t,\tilde z),z(\tilde t,\tilde z))}$ and ${\tilde
  N(\tilde t,\tilde z) = N(t(\tilde t,\tilde z),z(\tilde t,\tilde
  z))}$.
\\
As in \cite{Sc96vs} we map the coordinates $(\tilde t,\tilde z,\tilde
x,\tilde y)$ to $(t,z,x,y)$ by
\begin{mathletters}
\label{koorTrans}
\begin{eqnarray}
  \tilde t & = & 8 \> \frac{t^2+z^2}{\left(t^2-z^2\right)^2}, 
    \label{tTrans} \\
  \tilde z & = & 16 \> \frac{t\,z}{\left(t^2-z^2\right)^2}, 
    \label{zTrans} \\
  \tilde x & = & x, \label{xTrans} \\
  \tilde y & = & y. \label{yTrans}
\end{eqnarray}
\end{mathletters}
As we want to get all asymptotically A3 solutions within the metric
ansatz (\ref{physMetrik}), we have to know whether the coordinate
transformation (\ref{koorTrans}) for the ``compactification'' is essentially
unique. As there are two independent, non-vanishing, hypersurface-orthogonal
spacelike Killing vector fields the choice of null directions in the two
dimensional subspace orthogonal to the Killing orbits is unique up to
a null boost and those null directions cover the whole globally
hyperbolic part of the spacetime. The ``compactification'' $u=f(\tilde u)$
and $v=g(\tilde v)$ must lead to a regular and smooth unphysical
metric which fixes the ``compactification'' up to a
factor. Transformation (\ref{koorTrans}) is this essentially unique
``coordinate compactification''.
\\
The transformation (\ref{koorTrans}) maps $(\tilde t,\tilde z) \in
(\,]0,\infty[\,,\,]-t,t[\,)$ to $(t,z) \in
(\,]\infty,0[\,,\,]-t,t[\,)$. The notation $]\infty,0[$ emphases that
an increasing $\tilde t$ corresponds to a decreasing $t$. The inverse
of transformation (\ref{koorTrans}) is
\begin{mathletters}
\label{invKoorTrans}
\begin{eqnarray}
  t & = & \sqrt{\frac{2}{\tilde t - \tilde z}}
          + \sqrt{\frac{2}{\tilde t + \tilde z}}, 
    \label{invtTrans} \\
  z & = & \sqrt{\frac{2}{\tilde t - \tilde z}}
          - \sqrt{\frac{2}{\tilde t + \tilde z}}.
    \label{invzTrans}
\end{eqnarray}
\end{mathletters}
With the definitions
\begin{mathletters}
\label{unphysOmMetrik}
\begin{eqnarray}
  \Omega & = & \frac{1}{4} \left( t^2 - z^2 \right), 
    \label{unphysOm} \\
\noalign{and}
  g & = &
  \frac{4 \sqrt{2}}{\sqrt{t^2+z^2}} \> e^N \> ( - dt^2 + dz^2 )
  + \frac{1}{2} \left( t^2 + z^2 \right) 
    \left( e^W dx^2 + e^{-W} dy^2 \right) 
    \label{unphysg}
\end{eqnarray}
\end{mathletters}
we have $g = \Omega^2 \> \tilde g$. For $W$ and $N$ smooth on
$(t,z)=(\,]0,\infty[\,,\,[-t,t]\,)$ the spacetime $(\tilde M,\tilde
g)$ is by definition asymptotically A3. 
\\
The equations (\ref{physGl}) become
\begin{mathletters}
\label{unphysGl}
\begin{eqnarray}
  ( t^4 - z^4 ) \left( W_{,tt} - W_{,zz} \right)
    - 2 \, t \left( 3 \, z^2 + t^2 \right) W_{,t}
    - 2 \, z \left( z^2 + 3 \, t^2 \right) W_{,z}
  & = & 0, \label{unphysWllnGlW} \\
  N_{,t}
  + \frac{t^2+z^2}{4 \left(t^2-z^2\right)^2}
  \Biggm( t \> \left( 3\,z^2 + t^2 \right) 
          \left( \left(W_{,t}\right)^2 + \left(W_{,z}\right)^2
          \right)
  + 2 \, z \, \left( z^2 + 3 \, t^2 \right) 
    \, W_{,t} \, W_{,z}
  \Biggm)
  & = & 0, \label{unphystGlM} \\
\noalign{and}
  N_{,z}
  + \frac{t^2+z^2}{4 \left(t^2-z^2\right)^2}
  \Biggm( z \> \left( z^2 + 3\,t^2 \right) 
          \left( \left(W_{,t}\right)^2 + \left(W_{,z}\right)^2
          \right)
  + 2 \, t \, \left( 3\,z^2 + t^2 \right) 
    \, W_{,t} \, W_{,z}
  \Biggm)
  & = & 0. \label{unphysrGlM}
\end{eqnarray}
\end{mathletters}
These equations are singular on ${\cal J}$ where
$t^2-z^2=0$. Therefore giving smooth data $W$ and $W_{,t}$ on some
slice $(t_0,[-t_0,t_0])$ one does not know a priori whether the
solution $W$ remains
smooth. But for $W(t,z)$ smooth it follows from equation
(\ref{unphysWllnGlW}) that at $t=z$ the directional derivative
$W_{,t}+W_{,z}=0$ and at $t=-z$ the directional derivative
$W_{,t}-W_{,z}=0$. Therefore any smooth solution $W$ of
(\ref{unphysWllnGlW}) necessarily has the form 
\begin{equation}
\label{WA3}
  W(t,z) = f(t,z) {(t^2-z^2 )} + \const
\end{equation}
with smooth $f$. By a rescaling of $x$ and $y$
one can always achieve $\const=0$. The equations (\ref{unphysGl})
become
\begin{mathletters}
\label{unphysGlreg}
\begin{eqnarray}
  \left( f_{,tt} - f_{,zz} \right)
  - \frac{2 \, z}{t^2+z^2} f_{,z} + \frac{2 \, t}{t^2+z^2} f_{,t}
  & = & 0, \label{unphysWllnGlWreg} \\
  N_{,t}
  + \frac{t^2+z^2}{4}
  \Biggm( 4 \, t \left( f + 2  \, z  \, f_{,z} \right) f
          + t \left( 3 \, z^2 + t^2 \right) \left( f_{,z} \right)^2
          + 4  \, f \left( t^2 + z^2 \right) f_{,t} \hfill
          \qquad
          & & \nonumber \\
          {} + 2  \, z \left( z^2 + 3  \, t^2 \right) f_{,t}  \, f_{,z}
          + t \left( 3 z^2 + t^2 \right) \left( f_{,t} \right)^2
  \Biggm)
  & = & 0, \label{unphystGlMreg} \\
\noalign{and}
  N_{,z}
  + \frac{t^2+z^2}{4}
  \Biggm( 4  \, z \left( f + 2  \, t  \, f_{,t} \right) f
          + z \left( z^2 + 3  \, t^2 \right) \left( f_{,z} \right)^2
          + 4  \, f \left( t^2 + z^2 \right) f_{,z}
          \qquad
          \hfill & & \nonumber
          \\
          {} + 2  \, t \left( 3  \, z^2 + t^2 \right) f_{,t}  \, f_{,z}
          + z \left( z^2 + 3  \, t^2 \right) \left( f_{,t} \right)^2
  \Biggm)
  & = & 0, \label{unphysrGlMreg}
\end{eqnarray}
\end{mathletters}
which is regular for $t>0$.
\\
If we give smooth data $f$ and $f_{,t}$ on
$(t_0,\,{]-\infty,\infty[}\,)$ equation (\ref{unphysWllnGlWreg}) can be
solved on $(\,{]0,\infty[}\,,\,{]-\infty,\infty[}\,)$. The solution
$f(t,z)$ is smooth. Therefore integration of (\ref{unphystGlMreg}) and
(\ref{unphysrGlMreg}) gives a smooth $N(t,z)$. Hence we have
asymptotically A3 solutions of Einstein's equation.
\\
As the solutions in general do not have a timelike Killing vector
field, as do 
spherical symmetric vacuum solutions they are candidates for exact
solutions with gravitational radiation. The expression for the
Bondi mass written in terms of unphysical variables evaluated at the
conformal boundary is
\begin{eqnarray}
  \label{mBondi}
  m_{\mathrm Bondi} & = &
    \left. c \> \cdot \> e^{-N} \right|_{\cal J},
\end{eqnarray}
with a positive constant $c$ depending on the period of $x$ and
$y$ \cite{Fo97XX}. The derivatives along the null infinities are
decreasing, which can be seen by using equation
(\ref{unphystGlMreg}) and (\ref{unphysrGlMreg}).  
\\[\baselineskip]
Two everywhere regular solutions of (\ref{unphysWllnGlWreg}) can
easily be guessed. The first, $f(t,z) = c$, corresponds to 
\begin{mathletters}
\label{ersteLsg}
\begin{eqnarray}
  \tilde W(\tilde t,\tilde z) & = & 
  \frac{8 \, c}{\sqrt{{\tilde t}^2 - {\tilde z}^2}}, \label{ersteLsgtW}\\
  \tilde N(\tilde t,\tilde z) & = & 
  \frac{- 16 \, c^2 \, {\tilde t}^2}{\left(\tilde t^2 - \tilde z^2\right)^2},
  \label{ersteLsgtN}
\end{eqnarray}
\end{mathletters}
which is up to the constant $c$ the solution (3.6) of \cite{Sc96vs}. 
\\
The second, $f(t,z) = c\,t\,z$, corresponds to
\begin{mathletters}
\label{zweiteLsg}
\begin{eqnarray}
  \tilde W(\tilde t,\tilde z) & = & 
  \frac{32 \, c \, {\tilde z}}{{\sqrt{{\tilde t}^2 - {\tilde
          z}^2}}^{\,3}} , 
\label{zweiteLsgtW}\\
  \tilde N(\tilde t,\tilde z) & = & 
  \frac{- 128 \, c^2 \, {\tilde t}^2 \left( 8 \tilde z^2 + \tilde t^2 \right)}
       {\left(\tilde t^2 - \tilde z^2\right)^4}.
\label{zweiteLsgtN}
\end{eqnarray}
\end{mathletters}
\\[\baselineskip]
For general Gowdy models \cite{BeM93ni}, 
\begin{equation}
  \label{physMetrikunp}
  \tilde g = \frac{1}{\sqrt{\tilde t}} \> e^{\tilde N} 
               ( -d{\tilde t}^2 + d{\tilde z}^2 ) 
             + \tilde t \, 
               \left( e^{\tilde W} \, d{\tilde x}^2 
                      + 2 e^{\tilde W} \, d{\tilde x} \, d{\tilde y} 
                      + ( e^{\tilde W} {\tilde Q}^2 +
                           e^{-\tilde W} ) \, d{\tilde y}^2 \right),
\end{equation}
an equivalent procedure carries through. It is again necessary and
sufficient for asymptotically A3 that both of the functions ${\tilde
 W}$ and ${\tilde Q}$, which satisfy a system of nonlinear, coupled wave
equations, fall off such that
\begin{eqnarray}
\label{WA3unp}
  W(t,z) & = & f(t,z) {(t^2-z^2)} + \const \\
\noalign{and}
\label{QA3unp}
  Q(t,z) & = & g(t,z) {(t^2-z^2)} + \const.
\end{eqnarray}
The wave equations become regular equations for $f$ and $g$ and the
equations for $N$ are also regular. But due to the nonlinear coupling
of the wave equations, guessing solutions which are not polarized 
($Q=0$), may be difficult.
\\[\baselineskip]
It is a pleasure for me to thank Bernd Schmidt, Jim Isenberg, Vincent
Moncrief and Helmut Friedrich for helpful discussions.
\end{document}